\documentclass[a4paper,prl,showpacs,preprint]{revtex4}%
\usepackage{amsfonts}
\usepackage{amsmath}
\usepackage{amssymb}
\usepackage{graphicx}%
\begin{document}
\preprint{quant-ph/0308025}
\title{Engineering squeezed states in high-Q cavities\thanks{Published in Phys. Rev.
A \textbf{69}, 035802 (2004)}}
\author{N. G. de Almeida$^{\text{1,2}}$, R. M. Serra$^{\text{3,4}}$, C. J.
Villas-B\^{o}as$^{\text{4}}$, and M. H. Y. Moussa$^{\text{4}}$}
\affiliation{$^{1}$ Departamento de Matem\'{a}tica e F\'{\i}sica, Universidade Cat\'{o}lica
de Goi\'{a}s, P.O. Box 86, Goi\^{a}nia, 74605-010, Goi\'{a}s,
Brazil.\linebreak$^{2}$\textit{\ }Instituto de F\'{\i}sica, Universidade
Federal de Goi\'{a}s, Goi\^{a}nia, 74.001-970, Goi\'{a}s, Brazil.\linebreak%
$^{3}$ Optics Section, The Blackett Laboratory, Imperial College, London, SW7
2BZ, United Kingdom.\linebreak$^{4}$\textit{\ }Departamento de F\'{\i}sica,
Universidade Federal de S\~{a}o Carlos, P.O. Box 676, S\~{a}o Carlos,
13565-905, S\~{a}o Paulo, Brazil.}

\begin{abstract}
While it has been possible to build fields in high-Q cavities with a high
degree of squeezing for some years, the engineering of arbitrary squeezed
states in these cavities has only recently been addressed [Phys. Rev. A
\textbf{68}, 061801(R) (2003)]. The present work examines the question of
\ how to squeeze any given cavity-field state and, particularly, how to
generate the squeezed displaced number state and the squeezed macroscopic
quantum superposition in a high-$Q$ cavity.

Journal Ref.: Phys. Rev. A \textbf{69}, 035802 (2004)

\end{abstract}
\maketitle

Statistical properties of squeezed states of light have been widely
investigated and the possibility of applying squeezing properties to the
understanding of fundamental physical phenomena, as well as to solving
technological problems, has been recognized \cite{Dodonov}. As far as
fundamental phenomena are concerned, the antibunching or sub-Poissonian photon
statistic related to squeezed states has revealed unequivocal features of the
quantum nature of light \cite{Stoler}. In addition, squeezed-state
entanglements were recently employed for experimental demonstration of quantum
teleportation of optical coherent states \cite{Braunstein}. In technology, an
improvement of the signal-to-noise ratio in optical communication has been
proposed by reducing the quantum fluctuations in one quadrature component of
the field at the expense of the amplified fluctuations in another component
\cite{71}. Moreover, the possibility of using the quadrature component with
reduced quantum noise of a squeezed state as a pointer for the measurement of
weak signals has been suggested for the detection of gravitational waves as
well as for sensitive interferometric and spectroscopic measurements
\cite{Caves}.

Although squeezed light is mainly supplied by nonlinear optical media as
running waves, through backward \cite{98} or forward \cite{99} four-wave
mixing and parametric down-conversion \cite{100}, the dynamics of the
Jaynes-Cummings model (JCM) of atom-field interaction leads to standing
squeezed states of the electromagnetic field in cavity quantum electrodynamics
(QED) or the motional degree of freedom in ion traps \cite{Meystre}. Whereas
cavity-field squeezing in the JCM is rather modest, about 20\% for low average
photon number, squeezing of up to 75\% can be obtained with selective atomic
measurements \cite{Gerry}.{\large \ }However, such squeezed states (and those
obtained in the schemes employing atom-field interactions \cite{BW}) do have
not resulted from the unitary evolution $S(\xi)\left\vert \Psi\right\rangle $;
in other words, the experimenter is not able to squeeze any desired state
$\left\vert \Psi\right\rangle $ previously prepared in the cavity ($S(\xi)$
stands for the squeeze operator and $\xi$ for a set of group parameters). In
the present proposal we consider exactly the question of how to squeeze any
given cavity-field state $\left\vert \Psi\right\rangle $ and, in particular,
how to generate i) a squeezed displaced number state (SDNS) and ii) a squeezed
Schr\"{o}dinger-cat-like state (SSCS) in a high-Q cavity.

The SDNS, $\left\vert \xi;\alpha;n\right\rangle $, is obtained by the action
of the displacement operator $D(\alpha)=\exp[\frac{1}{2}\left(  \alpha^{\ast
}a-\alpha a^{\dagger}\right)  ]$, followed by the squeeze operator
$S(\xi)=\exp\left[  \frac{1}{2}\left(  \xi^{\ast}a^{2}-\xi a^{\dagger
2}\right)  \right]  $, on the number state $\left\vert \xi;\alpha
;n\right\rangle \equiv S(\xi)D(\alpha)\left\vert n\right\rangle $. It is
readily seen that the SDNS contains various special cases such as the number
state ($\xi=\alpha=0$), coherent state ($\xi=n=0$), squeezed number state
($\alpha=0$), displaced number state ($\xi=0$), and so on. Therefore, the SDNS
allows a unified approach incorporating all these states, and their
properties. Although the statistical properties of the SDNS are well known
\cite{Kral}, the generation of SDNS in a cavity has not been reported yet.
Recently, we showed how to achieve an effective quadratic Hamiltonian leading
to the parametric frequency conversion process in cavity QED \cite{Celso},
opening the way for generation of a cavity SDNS.

To avoid experimental complications stemming from introducing a nonlinear
crystal inside a cavity, the squeeze operator is built from the dispersive
interaction of the cavity mode with a driven three-level atom \cite{Celso}. As
sketched in Fig. 1, the atomic system is in the ladder configuration, where an
intermediate atomic level ($\left\vert i\right\rangle $) lies between the
ground ($\left\vert g\right\rangle $) and excited ($\left\vert e\right\rangle
$) states. The quantized cavity mode of frequency $\omega$ couples
dispersively both transitions $\left\vert g\right\rangle $ $\leftrightarrow$
$\left\vert i\right\rangle $ and $\left\vert e\right\rangle $ $\leftrightarrow
$ $\left\vert i\right\rangle $, with coupling constants $\lambda_{g}$ and
$\lambda_{e}$, respectively, and detuning $\delta=\left\vert \omega
-\omega_{\ell i}\right\vert $ ($\ell=g,e$). A classical field of frequency
$\omega_{0}=2\omega+\Delta$ drives  the atomic transition $\left\vert
g\right\rangle $ $\leftrightarrow$ $\left\vert e\right\rangle $ dispersively,
with coupling constant $\Omega$. The transition $\left\vert g\right\rangle $
$\leftrightarrow$ $\left\vert e\right\rangle $ may be induced by applying a
sufficiently strong electric field. While the quantum field promotes a
two-photon interchange process, the classical driving field constitutes the
source of the parametric amplification.

The Hamiltonian of our model, under the rotating wave approximation, is given
by $H=H_{0}+V$, where
\begin{subequations}
\begin{align}
H_{0} &  =\hbar\omega a^{\dagger}a-\hbar\omega\left\vert g\right\rangle
\left\langle g\right\vert +\hbar\delta\left\vert i\right\rangle \left\langle
i\right\vert +\hbar\omega\left\vert e\right\rangle \left\langle e\right\vert
,\label{Eq1a}\\
V &  =\hbar\left(  \lambda_{g}a\left\vert i\right\rangle \left\langle
g\right\vert +H.c.\right)  +\hbar\left(  \lambda_{e}a\left\vert e\right\rangle
\left\langle i\right\vert +\mathrm{H.c}\right)  \nonumber\\
&  +\hbar\left(  \Omega\left\vert e\right\rangle \left\langle g\right\vert
e^{-i\omega_{0}t}+H.c.\right)  ,\label{Eq1b}%
\end{align}
with $a^{\dagger}$ ($a$) standing for the creation (annihilation) operator of
the quantized cavity mode. Writing $H$ in the interaction picture [through the
unitary transformation $U_{0}=\exp\left(  -iH_{0}t/\hbar\right)  $] and then
applying the transformation\ $U=\exp\left[  -i\delta t\left(  \left\vert
g\right\rangle \left\langle g\right\vert +\left\vert e\right\rangle
\left\langle e\right\vert \right)  \right]  $, we obtain the Hamiltonian
$\mathcal{H}=U_{0}^{\dagger}U^{\dagger}HUU_{0}-H_{0}-\hbar\delta\left(
\left\vert g\right\rangle \left\langle g\right\vert +\left\vert e\right\rangle
\left\langle e\right\vert \right)  $. If the dispersive transitions are
sufficiently{\large \ }detuned, i.e., $\delta\gg$ $\left\vert \lambda
_{g}\right\vert $,$\left\vert \lambda_{e}\right\vert $,$\left\vert
\Omega\right\vert $, we obtain the adiabatic solutions for the transition
operators $\sigma_{ig}$ and $\sigma_{ei}$ ($\sigma_{kl}\equiv\left\vert
k\right\rangle \left\langle l\right\vert $, $k,l=g,i,e$.) by setting
$d\sigma_{ig}/dt=d\sigma_{ei}/dt=0$; solving the resulting system, and
inserting these adiabatic solutions for $\sigma_{ig}$ and $\sigma_{ei}$ into
$\mathcal{H}$ \ (for more details see \cite{Celso}), the Hamiltonian becomes
\end{subequations}
\begin{align}
\mathcal{H} &  \approx-\hbar\delta\left(  \sigma_{gg}+\sigma_{ee}\right)
+\hbar\left(  \Omega\operatorname*{e}\nolimits^{-i\Delta t}\sigma
_{eg}+H.c\right)  -\frac{\hbar}{\delta}\left\{  \left(  2a^{\dagger
}a+1\right)  \right.  \nonumber\\
&  \left.  \times\left[  \left\vert \lambda_{g}\right\vert ^{2}\sigma
_{gg}-\left(  \left\vert \lambda_{g}\right\vert ^{2}+\left\vert \lambda
_{e}\right\vert ^{2}\right)  \sigma_{ii}+\left\vert \lambda_{e}\right\vert
^{2}\sigma_{ee}+\frac{\left\vert \lambda_{g}\right\vert ^{2}+\left\vert
\lambda_{e}\right\vert ^{2}}{2\delta}\left(  \Omega\operatorname*{e}%
\nolimits^{-i\Delta t}\sigma_{eg}+\mathrm{H.c.}\right)  \right]  \right\}
\nonumber\\
&  -\frac{\hbar}{\delta}\left\{  2\left(  \lambda_{g}\lambda_{e}a^{2}%
\sigma_{eg}+\mathrm{H.c.}\right)  +\frac{1}{\delta}\left(  \lambda_{g}%
\lambda_{e}\Omega^{\ast}\operatorname*{e}\nolimits^{i\Delta t}a^{2}%
+\mathrm{H.c.}\right)  \left(  \sigma_{gg}+\sigma_{ee}-2\sigma_{ii}\right)
\right\}  \label{Eq2}%
\end{align}
The state vector associated with the Hamiltonian (\ref{Eq2}) can be written
using
\begin{equation}
|\Psi\left(  t\right)  \rangle=\left\vert g\right\rangle \left\vert \Phi
_{g}\left(  t\right)  \right\rangle +\left\vert i\right\rangle \left\vert
\Phi_{i}\left(  t\right)  \right\rangle +\left\vert e\right\rangle \left\vert
\Phi_{e}\left(  t\right)  \right\rangle \mathrm{{,}}\label{Eq3}%
\end{equation}
where $|\Phi_{\ell}\left(  t\right)  \rangle=\int\frac{d^{2}\alpha}{\pi
}\mathcal{A}_{\ell}\left(  \alpha,t\right)  |\alpha\rangle$ for $\ell=g,i,e$,
the complex quantity $\alpha$ standing for the eigenvalues of $a$, and
$\mathcal{A}_{\ell}\left(  \alpha,t\right)  =\left\langle \alpha
,\ell\left\vert \Psi\left(  t\right)  \right.  \right\rangle $ represents the
set of expansion coefficients for $|\Phi_{\ell}\left(  t\right)  \rangle$ in
the basis of coherent states, $\left\{  |\alpha\rangle\right\}  $. Using the
orthogonality of the atomic states and Eqs. (\ref{Eq2}) and (\ref{Eq3}) we
obtain the uncoupled time-dependent (TD) Schr\"{o}dinger equation for the
atomic subspace $\left\vert i\right\rangle $ (in the Schr\"{o}dinger
picture):
\begin{equation}
i\hbar\frac{d}{dt}|\Phi_{i}\left(  t\right)  \rangle=\mathcal{H}_{i}|\Phi
_{i}\left(  t\right)  \rangle\mathrm{{,}}\label{Eq4}%
\end{equation}%
\begin{equation}
\mathcal{H}_{i}=\hbar\varpi a^{\dagger}a+\hbar\left(  \xi\operatorname*{e}%
\nolimits^{-i\nu t}a^{\dagger^{2}}+\xi^{\ast}\operatorname*{e}\nolimits^{i\nu
t}a^{2}\right)  \label{Eq5}%
\end{equation}
where $\varpi=\omega+\chi$ $\left(  \chi=\left.  2\left(  \left\vert
\lambda_{g}\right\vert ^{2}+\left\vert \lambda_{e}\right\vert ^{2}\right)
\right/  \delta\right)  $ stands for the effective frequency of the cavity
mode, while $\xi=2\Omega\lambda_{g}^{\ast}\lambda_{e}^{\ast}/\delta
^{2}=\left\vert \xi\right\vert \operatorname*{e}\nolimits^{-i\Theta}$ and
$\nu=2\omega+\Delta$ are the effective amplitude and frequency of the
parametric amplification field. For subspace $\left\{  \left\vert
g\right\rangle ,\left\vert e\right\rangle \right\}  $ there is a TD
Schr\"{o}dinger equation which couples the fundamental and excited atomic
states. Therefore, when we initially prepare the atom in the intermediate
level $\left\vert i\right\rangle $, the dynamics of the atom-field dispersive
interactions, governed by the effective Hamiltonian (\ref{Eq5}), results in a
cavity mode with shifted frequency submitted to a parametric amplification process.

For the present purpose we consider the resonant regime, where the classical
driving field has the same frequency $\varpi$ as the effective cavity mode, so
that $\nu=2\varpi$ (i.e. $\Delta=2\chi$). (A treatment of the off-resonant
interaction between the effective cavity mode and the driving field was
investigated in Ref.\cite{Celso}.) The evolution of the cavity field state, in
the interaction picture, is governed by a squeeze operator such as $|\Phi
_{i}\left(  t\right)  \rangle=S(\xi,t)|\Phi_{i}\left(  t_{0}\right)  \rangle$,
where
\begin{equation}
S(\xi,t)=\exp\left[  -i\left(  \xi a^{\dagger2}+\xi^{\ast}a^{2}\right)
t\right]  . \label{Eq6}%
\end{equation}
The degree of squeezing in the resonant regime is determined by the factor
$r(t)=2\left\vert \xi\right\vert t$, while the squeeze angle is given by
$\varphi=\pi/2-\Theta$. For a specific cavity mode and atomic system, the
parameter $r(t)$ can be adjusted in accordance with the coupling strength
$\Omega$ and the interaction time $t$.

With this squeeze operator in hand, we are able to show how to engineer the
two specific squeezed states already mentioned: i) the SDNS $\left\vert
\xi;\alpha;n\right\rangle \equiv S(\xi)D(\alpha)\left\vert n\right\rangle $
and ii) the SSCS $S(\xi)\left[  \mathcal{N}\left(  \left\vert \alpha
\right\rangle +e^{i\phi}\left\vert -\alpha\right\rangle \right)  \right]  $
($\mathcal{N}$ being the normalization factor).

i) Starting with the SDNS, the first step is to prepare the cavity field in
the Fock state $\left\vert n\right\rangle $, which can in principle be done by
any of the proposals in Ref. \cite{EngFock}. However, we observe that multiple
of $2$ number states $\left\vert n=2m\right\rangle $ ($m=1,2,..$) can be
generated as a by-product of the present scheme with the driving field
switched off. In fact, considering $\Omega=0$ and disregarding state
$\left\vert i\right\rangle $, the Hamiltonian (\ref{Eq2}) becomes
\begin{align}
\widetilde{\mathcal{H}}  &  =-\hbar\left[  \left(  \delta+\frac{\left\vert
\lambda_{g}\right\vert ^{2}}{\delta}\right)  +\frac{2\left\vert \lambda
_{g}\right\vert ^{2}}{\delta}a^{\dagger}a\right]  \sigma_{gg}-\hbar\left[
\left(  \delta+\frac{\left\vert \lambda_{e}\right\vert ^{2}}{\delta}\right)
+\frac{2\left\vert \lambda_{e}\right\vert ^{2}}{\delta}a^{\dagger}a\right]
\sigma_{ee}\nonumber\\
&  +2\frac{\hbar}{\delta}\left(  \lambda_{g}\lambda_{e}a^{2}\sigma
_{eg}+H.c.\right)  . \label{Eq7}%
\end{align}
This Hamiltonian allows the transition\textbf{\ }$\left\vert n,e\right\rangle
\longleftrightarrow\left\vert n+2,g\right\rangle $. Assuming that the atom is
prepared in the state $\left\vert e\right\rangle $, we obtain the following
evolution
\begin{align}
e^{-i\widetilde{\mathcal{H}}t/\hbar}\left\vert n,e\right\rangle  &
=\left\vert \Upsilon\right\vert ^{2}\left[  \frac{e^{-it\eta_{+}}}{\left\vert
\Upsilon\right\vert ^{2}+(\eta_{+}-\Lambda)^{2}}+\frac{e^{-it\eta_{-}}%
}{\left\vert \Upsilon\right\vert ^{2}+(\eta_{-}-\Lambda)^{2}}\right]
\left\vert n,e\right\rangle \nonumber\\
&  +\Upsilon^{\ast}\left[  \frac{(\eta_{+}-\Lambda)^{2}e^{-it\eta_{+}}%
}{\left\vert \Upsilon\right\vert ^{2}+(\eta_{+}-\Lambda)^{2}}+\frac{(\eta
_{-}-\Lambda)^{2}e^{-it\eta_{-}}}{\left\vert \Upsilon\right\vert ^{2}%
+(\eta_{-}-\Lambda)^{2}}\right]  \left\vert n+2,g\right\rangle , \label{Eq88}%
\end{align}
where%
\begin{subequations}
\begin{align}
\eta_{\pm}  &  =\frac{1}{2}\left(  \Lambda+\Xi\pm\sqrt{\left(  \Lambda
-\Xi\right)  ^{2}+4\Upsilon^{2}}\right)  ,\\
\Lambda &  =\delta+\frac{\left\vert \lambda_{e}\right\vert ^{2}}{\delta}%
+\frac{2\left\vert \lambda_{e}\right\vert ^{2}}{\delta}n,\\
\Xi &  =\delta+\frac{\left\vert \lambda_{g}\right\vert ^{2}}{\delta}%
+\frac{2\left\vert \lambda_{g}\right\vert ^{2}}{\delta}\left(  n+2\right)  ,\\
\Upsilon &  =\frac{2\lambda_{g}\lambda_{e}}{\delta}\sqrt{\left(  n+2\right)
\left(  n+1\right)  }.
\end{align}
If the initial cavity state is $\left\vert n\right\rangle $, the probability
of detecting the atomic level $\left\vert g\right\rangle $ is given by
$P_{g,n}(t)=\left\Vert \left\langle g\right\vert e^{-i\widetilde{\mathcal{H}%
}t/\hbar}\left\vert n,e\right\rangle \right\Vert ^{2}$. In order to prepare
the state $\left\vert n+2\right\rangle $, the atom-field interaction time must
be adjusted to maximize $P_{g,n}$, that is $t=\pi\left(  \eta_{+}-\eta
_{-}\right)  ^{-1}$, and the success probability is given by
\end{subequations}
\begin{align*}
P_{g},_{n}  &  =\left\vert \Upsilon\right\vert ^{2}\left\{  \left(
\frac{(\eta_{+}-\Lambda)^{2}}{\left\vert \Upsilon\right\vert ^{2}+(\eta
_{+}-\Lambda)^{2}}\right)  ^{2}+\left(  \frac{(\eta_{-}-\Lambda)^{2}%
}{\left\vert \Upsilon\right\vert ^{2}+(\eta_{-}-\Lambda)^{2}}\right)
^{2}\right. \\
&  \left.  -\frac{(\eta_{-}-\Lambda)^{2}(\eta_{+}-\Lambda)^{2}}{\left[
\left\vert \Upsilon\right\vert ^{2}+(\eta_{-}-\Lambda)^{2}\right]  \left[
\left\vert \Upsilon\right\vert ^{2}+(\eta_{+}-\Lambda)^{2}\right]  }\right\}
.
\end{align*}

Therefore, starting with an empty cavity and passing a stream of $m$
three-level atoms through it with an adequately adjusted interaction time for
each atom $t_{k}=\pi\left(  \eta_{+}-\eta_{-}\right)  ^{-1}$ (where the
subscript $k$ indicates the $k$th atom), we have a probabilistic technique for
building multiple of 2 number states $\left\vert n=2m\right\rangle $. Each
atom is supposed to be detected in the state $\left\vert g\right\rangle $
before the subsequent atom enters the cavity.

To illustrate this technique, we will consider typical parameter values which
follow from Rydberg-states where the intermediate state $\left\vert
i\right\rangle $ (an $(n-1)P_{3/2}$ level) is nearly halfway between
$\left\vert g\right\rangle $ (an $(n-1)S_{1/2}$ level) and $\left\vert
e\right\rangle $ (an $nS_{1/2}$ level), namely $\left\vert \lambda
_{g}\right\vert \sim\left\vert \lambda_{e}\right\vert \sim7\times10^{5}%
$s$^{-1}$ \cite{BRH}, and we will assume the detuning $\left\vert
\delta\right\vert \sim1\times10^{7}$s$^{-1}$. We show in Table I the
interaction time and the probability of successfully building the states
$\left\vert 2\right\rangle ,\left\vert 4\right\rangle $ and $\left\vert
6\right\rangle $ by passing $1,2,$ and $3$ atoms, respectively, through the cavity.

After preparing the initial number state, the displacement operator is
implemented by connecting a microwave source to the cavity \cite{Raymond}. The
prepared cavity field $\left\vert n\right\rangle $ is displaced when the
microwave source is turning on, the amount of displacement being adjusted by
varying the time interval of injection of the classical microwave field.
Finally, a driven three-level atom (in this step $\Omega\neq0$) prepared in
the intermediate state $\left\vert i\right\rangle $ is sent through the cavity
to accomplish the squeezing operation. Particular cases of the SDNS, such as
the squeezed vacuum, $\left\vert \xi;0;0\right\rangle \equiv S(\xi)\left\vert
0\right\rangle $, or the squeezed coherent state (SCS), $\left\vert \xi
;\alpha;0\right\rangle \equiv S(\xi)\left\vert \alpha\right\rangle $, are
easily engineered by sending just one driven three-level atom through a cavity
initially prepared in the vacuum or the coherent state, respectively. The
degree of squeezing will be discussed further.

ii) As a second application, we consider the squeezed Schr\"{o}dinger-cat-like
state, SSCS. The Schr\"{o}dinger-cat-like state $\left\vert \Psi\right\rangle
$ is easily generated by sending a two-level atom across a cavity (initially
in a prepared coherent state $|\beta\rangle$) sandwiched by two Ramsey zones,
as reported in \cite{Brune}. Properly adjusting  the atom-field interaction in
the Ramsey zones and in the cavity, it is possible to obtain the state
\cite{Brune}%
\begin{equation}
\left\vert \Psi^{\pm}\right\rangle =\mathcal{N}_{\pm}\left(  c_{g}%
|\alpha\rangle\pm c_{e}|-\alpha\rangle\right)  \text{,}\label{Eq8}%
\end{equation}
where $\alpha=i\beta$, $\mathcal{N}_{\pm}$ is the normalization factor and the
$+$ ($-$) sign occurs if the atom is detected in state $|g\rangle$
($|e\rangle$). Finally, following the scheme proposed here, the SSCS is
achieved by sending a driven three-level atom through the cavity.

Let us discuss briefly the degree of squeezing achievable with the present
proposal, focussing our attention on the SCS and on the squeezed number state.
For an initial coherent state prepared in the cavity, the variance of the
squeezed quadrature is given by $\ \Delta X=\operatorname*{e}\nolimits^{-2r}%
/4$ \cite{Scully}. Assuming typical parameter values (cited above), and the
coupling strength $\Omega\sim7\times10^{5}$s$^{-1}$, we obtain $\left\vert
\xi\right\vert \sim6.8\times10^{3}$s$^{-1}$. For an atom-field interaction
time about $t\sim10^{-4}$s (which is one order of magnitude smaller than the
usual decay time of the open cavities used in experiments \cite{Haroche}) we
get the squeezing factor $r(t)\sim1.36$, resulting in a variance in the
squeezed quadrature of $\Delta X\sim1.6\times10^{-2}$ for the SCS. This
represents a high degree of squeezing, around 93\% with the passage of just
one driven three level atom. For the squeezed number state ---considering the
above parameters and an initial number state $n=2$, which can be generated
with the passage of just one atom through the cavity, as described above--- we
obtain a variance in the squeezed quadrature $\Delta X\sim8\times10^{-2}$,
representing a degree of squeezing around 67\%.

We note that for weakly-damped systems, such as fields trapped in realistic
high-$Q$ cavities, the lifetime of the squeezing is of the order of the
relaxation time of the cavity, a result which is valid even at absolute zero
\cite{Grabert}. Therefore, the dissipative mechanism of the cavity plays a
much milder role in the lifetime of the squeezing than in decoherence
phenomena \cite{LD}. Regarding atomic decay, note that for Rydberg levels the
damping effects can be safely neglected on typical interaction time scales. A
straightforward estimate of the fidelity of the prepared states under the
damping effects can be made through the phenomenological operator technique,
as described in \cite{NPR}.

In conclusion, we have shown how to engineer some squeezed states of the
radiation field in cavity QED, based on the interaction of the field with a
driven three-level atom. Particular states, such as the squeezed vacuum and
the squeezed coherent state, are easily engineered by sending just one atom
through the cavity, making our proposal attractive for experimental
implementation. To build the SDNS, an intermediate step is needed to prepare
the number state, as described previously, which clearly makes the SDNS less
attractive for experimental implementation. The SSCS is accomplished by
sending two atoms through the cavity, the first (a two-level atom interacting
dispersively with the cavity mode) to prepare the Schr\"{o}dinger-cat state,
as in Ref. \cite{Brune} and the second (as shown above) to execute the
squeezing operation. Finally, we would like to underline that up to 93\%
degree of squeezing of a field state, initially prepared in the coherent
state, may be achieved by passing a single three-level atom through the
cavity. This high degree of squeezing is crucial to the building of truly
mesoscopic superpositions with a large average photon number and also a large
\textquotedblleft distance\textquotedblright\ in phase space between the
centers of the quasi-probability distribution of the individual states
composing the prepared superposition \cite{Celso2}.

\begin{acknowledgments}
We wish to express our thanks for the support of FAPESP (under contracts
\#99/11617-0, \#00/15084-5, and \#02/02633-6) and CNPq (Instituto do
Mil\^{e}nio de Informa\c{c}\~{a}o Qu\^{a}ntica), Brazilian research funding agencies.
\end{acknowledgments}

\textbf{Figure Captions}

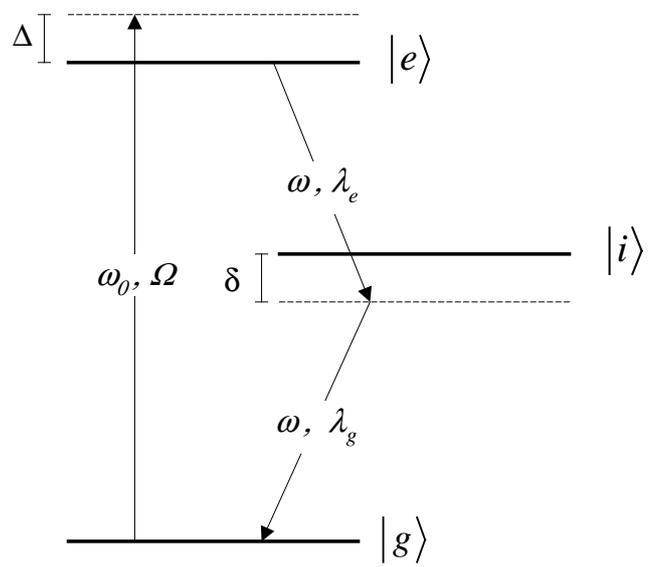
Fig. 1. Energy-level diagram of the three-level atom for the parametric
amplification scheme.

\newpage

\textbf{Tables}

Table I. Interaction time ($t_{k}$) and the probability ($P_{g,k}$) of
detecting the state $\left\vert g\right\rangle $ to engineer the number state
$\left\vert n=2m\right\rangle $ when $m$ atoms are passed through the cavity.
The total probability of success in building the state $\left\vert
n=2m\right\rangle $ is $\mathcal{P}=\prod_{k=1}^{m}P_{g,k}$.

\begin{center}%
\begin{tabular}
[c]{llll}\hline\hline
$\quad m~~$ & $\quad n~~~$ & $~~t_{k}$ $(10^{-5}s)$ & $\quad P_{g,m}$\\\hline
\end{tabular}

\begin{tabular}
[c]{cccc}%
$\quad1\quad$ & $\quad2\quad$ & $\quad0.9254\quad$ & $\quad~0.6$\\
$2$ & $4$ & $\quad0.4446\quad$ & $\quad~0.8$\\
$3$ & $6$ & $\quad0.2879\quad$ & $\quad~0.9$\\\hline\hline
\end{tabular}

\end{center}

\end{document}